# Nanodevices engineering and spin transport properties of MnBi$_2$Te$_4$ monolayer

Yipeng An[1]*, Kun Wang[2], Shijing Gong[3], Yusheng Hou[4,5], Chunlan Ma[6], Mingfu Zhu[7], Chuanxi Zhao[8], Tianxing Wang[1], Shuhong Ma[1], Heyan Wang[1], Ruqian Wu[4]* and Wuming Liu[9]*

**ABSTRACT:**

Two-dimensional (2D) magnetic materials are essential for the development of the next-generation spintronic technologies. Recently, layered van der Waals (vdW) compound MnBi$_2$Te$_4$ (MBT) has attracted great interest, and its 2D structure has been reported to host coexisting magnetism and topology. Here, we design several conceptual nanodevices based on MBT monolayer (MBT-ML) and reveal their spin-dependent transport properties by means of the first-principles calculations. The *pn*-junction diodes and sub-3-nm *pin*-junction field-effect transistors (FETs) show a strong rectifying effect and a spin filtering effect, with an ideality factor *n* close to 1 even at a reasonably high temperature. In addition, the *pip*- and *nin*-junction FETs give an interesting negative differential resistive (NDR) effect. The gate voltages can tune currents through these FETs in a large range. Furthermore, the MBT-ML has a strong response to light. Our results uncover the multifunctional nature of MBT-ML, pave the road for its applications in diverse next-generation semiconductor spin electric devices.

[1] School of Physics & Henan Key Laboratory of Boron Chemistry and Advanced Energy Materials, Henan Normal University, Xinxiang 453007, China
[2] Department of Physics and Astronomy, Mississippi State University, Mississippi State, MS 39762, USA; Department of Chemistry, Mississippi State University, Mississippi State, MS 39762, USA
[3] Key Laboratory of Polar Materials and Devices (MOE) & Department of Optoelectronics, East China Normal University, Shanghai 200062, China
[4] Department of Physics and Astronomy, University of California, Irvine, California 92697, USA
[5] School of Physics, Sun Yat-Sen University, Guangzhou 510275, China
[6] School of Physics and Technology, Suzhou University of Science and Technology, Suzhou, Jiangsu 215009, China
[7] Hebi National Lighting Co. Ltd., Hebi 458000, China
[8] Siyuan Laboratory, Guangdong Provincial Engineering Technology Research Center of Vacuum Coating Technologies and New Energy Materials, Department of Physics, Jinan University, Guangzhou 510632, China
[9] Beijing National Laboratory for Condensed Matter Physics, Institute of Physics, Chinese Academy of Sciences, Beijing 100190, China
*e-mail: ypan@htu.edu.cn; wur@uci.edu; wliu@iphy.ac.cn

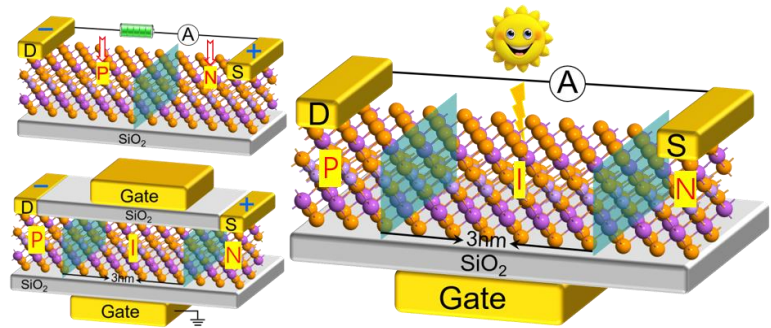

**INTRODUCTION**

The discovery of magnetic van der Waals (vdW) layered materials has inspired tremendous research interest recently as they provide new opportunities for the development of magnetic nanodevices. Scientists have utilized their spin degree of freedom to demonstrate exotic phenomena and devoted significant efforts to fabricate vdW ultrathin films down to the monolayer (ML) limit. It has been reported that the magnetic properties of vdW layered materials change significantly as their thickness reduces to atomically thin, and many of 2D vdW monolayers appear to be promising for spintronics applications.[1,2] In particular, enormous opportunities arise as a consequence of reduced dimensionality, which allows gating capabilities to tune their physical properties in a large range. Therefore, searching for functional 2D magnetic monolayers and designing conceptual nanodevices are extremely active and may forcefully drive the new technological development.

Several intrinsically magnetic 2D materials have attracted considerable attention in recent years, including $VX_2$ (X= S, Se, Te),[3-9] $CrGeTe_3$,[10-13] $Fe_3GeTe_2$,[14-17] and $CrX_3$(X = Cl, Br, I).[18-24] Furthermore, the discovery of $MnBi_2Te_4$(MBT) represents another step forward, as it integrates topological feature and alternating ferromagnetic (FM) or antiferromagnetic (AFM) ordering in a single system.[1] This renders MBT an ideal system for creating nanodevices with unprecedented magnetic and spintronic properties and, therefore, the system has already drawn extensive theoretical and experiment studies.[25-32] For instance, ultrathin MBT films were recently fabricated and their spin transport was successfully modulated by gate electrodes.[33,34] Despite substantial progresses, in-depth explorations of using MBT monolayers in diverse nanodevices are highly desired.

Here, we propose several conceptual MBT-based nanodevices and investigate their spin-dependent transport properties with the first-principles approach. We first construct *pn*-junction diodes with a MBT-ML and demonstrate its spin-dependent transport function. This incites us to build up several MBT-ML field-effect transistors that can be used for the design of spin logic circuits and storage units. We further propose MBT-based phototransistors and show the potential utilizations of MBT-ML in photoelectric nanodevices.

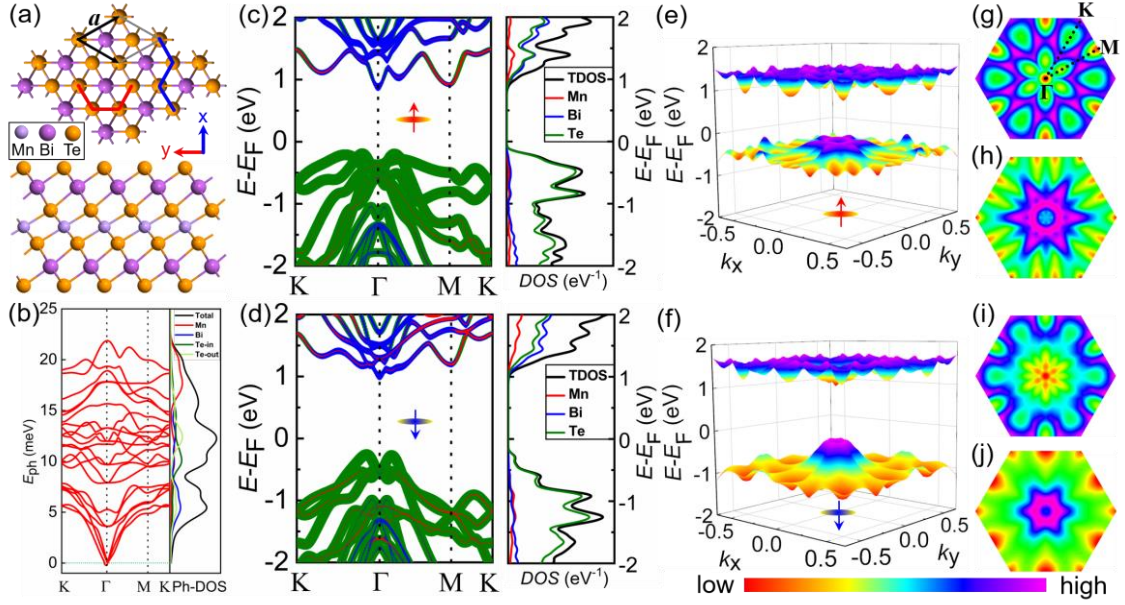

**Fig. 1 Geometric and electronic structures of MnBi$_2$Te$_4$ monolayer. a** Schematics of the top (upper) and side (lower) views of MnBi$_2$Te$_4$ monolayer. *x* and *y* axis are for zigzag and armchair directions, respectively. **b** Phonon band and projected phonon density of states. Element-projected electronic band and density of states for the spin-up (**c**) and -down (**d**) states. The $E_F$ is set to zero. **e** – **j** 3D views of the spin-resolved conduction and valence bands around the Γ point, as well as their 2D projections in the first Brillouin zone. The rainbow colormap shows the eigenvalues of conduction and valence bands from low (red) to high (purple).

## RESULTS AND DISCUSSION

**Atomic Structure and Electronic Properties of MnBi$_2$Te$_4$ Monolayer.** The ground-state atomic and electronic structures of MBT-ML are shown in Fig. 1. MBT-ML (Fig. 1a) has the $P\bar{3}m1$ space group (No. 164) and the optimized in-plane lattice parameter ***a*** of the hexagonal septuple-layer (SL) structure is 4.29 Å. The magnetic moment is 5 $\mu_B$ per unit cell, mostly contributed by Mn atoms, showing a high spin configuration *S* = 5/2, consistent with previous work.[35,36] MBT-ML was prepared on the Si (111) substrate in a so-called

SL-by-SL manner by alternative growing quintuple-layer of $Bi_2Te_3$ and bilayer of MnTe in molecular beam epitaxy.[33] The dynamic stability of the free-standing MBT-ML is confirmed by its phonon spectrum (Fig. 1b), which has no imaginary phonon modes. According to the projected phonon density of states (Ph-DOS) in Fig. 1b, the acoustic phonon branches stem from vibrations of Bi atoms, while the high- and low-frequency optical branches are from motions of central Mn and Te atoms, respectively.

Figure 1c, d show the element-projected spin-up and -down(dn) electronic bands and density of states (DOS), respectively. The MBT-ML has an indirect bandgap for either the spin-up ($E_g$ = 1.1 eV) or the spin-down ($E_g$ = 1.3 eV) state. The conduction bands are mainly contributed by the outer Bi and Te atoms, while the valence bands are mostly contributed by the Te atoms. The conduction-band minimum (CBM) is located at the Γ point, and the valence-band maximum (VBM) is between the Γ and K points. To give a more in-depth view of the band-dispersion relation around the Fermi level ($E_F$), we show the three-dimensional (3D) views of the conduction and valence bands of MBT-ML for the spin-up and -down states in Fig. 1e, f, respectively. Each energy-dispersion plot in the 2D *k*-space is hexagonal around the Γ point, which is more clearly illustrated in their projections in the first Brillouin zone (Fig. 1g:j).

The cone-shaped conduction-band near the Γ point (Fig. 1g, i) suggests mostly isotropic effective mass around the Γ point. They are 0.10 and 0.13 $m_e$ ($m_e$ is the free electron mass) for the spin-up and -down states, corresponding to Fermi velocities of 6.5 × 10$^5$ and 4.0 ×10$^5$ m/s, respectively, which is close to those of graphene and silicene.[37] The MBT-ML hence has higher carrier mobility and larger transmission possibility in the spin-up channel, and a spin-polarized electronic transport behavior is expected. Besides, both the energy dispersions and effective mass become anisotropic away from Γ point, such as between the *x* axis (along zigzag direction) and y axis (along armchair direction) shown in Fig. 1a, which should lead to anisotropic transport properties along these two directions like many other 2D materials.[6,38-40] Note that the bandgap becomes smaller (see Supplementary Fig. 1a) as considering the spin-orbit coupling (SOC), consistent with the previous report.[41] This suggests that the semiconductor nanodevices based on the MBT-ML is easily achieved when the SOC is included. In the following, we design several conceptual nanodevices of MBT-ML and mainly focus on its spin-polarized transport behaviors without SOC, which qualitatively retains its electronic transport properties based on a test (see Supplementary Fig. 1b).

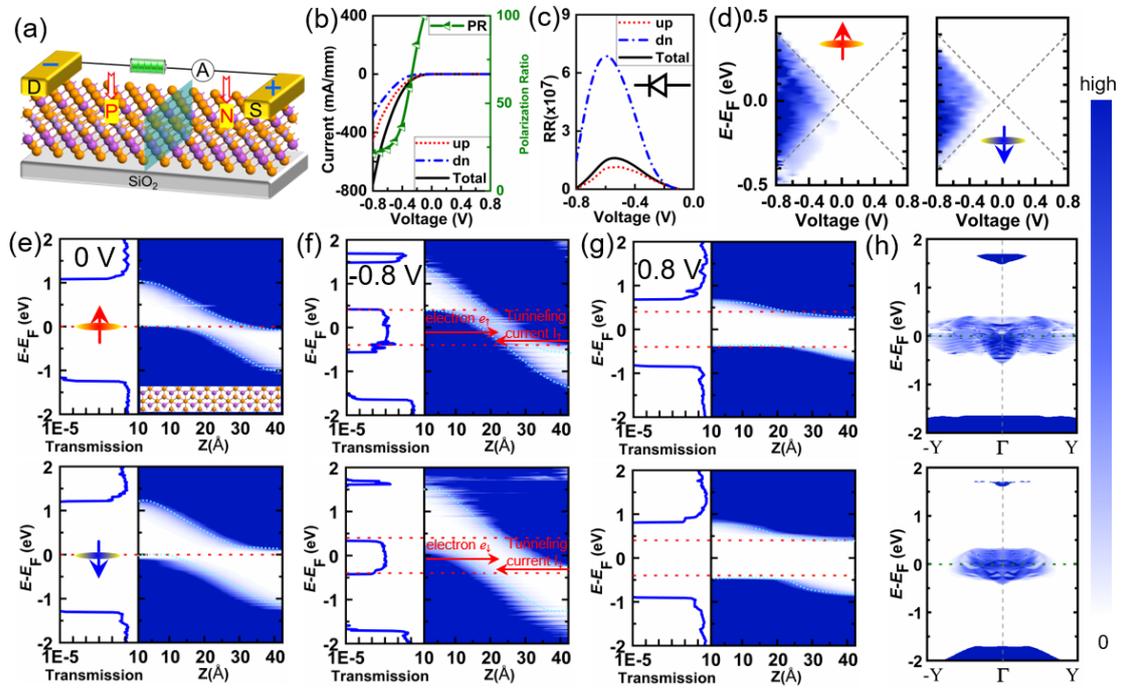

**Fig. 2 Transport properties of *pn*-junction diodes of MnBi$_2$Te$_4$ monolayer. a** Schematic of *pn*-junction diode of MnBi$_2$Te$_4$ monolayer. **b** Bias-dependent current and polarization ratio curves of the Z-type *pn*-junction diode of MBT-ML, as well as its rectifying ratio in (**c**). **d** Transmission spectra for the spin-up and -down states. **e – g** Spin-resolved transmission coefficient *T*(E) and projected local density of states under the biases of 0, –0.8, and 0.8 V. **h** *k*-dependent transmission coefficients *T*(E, k) at –0.8 V. The colormap shows the data of **d** to **h** from 0 (white) to high (blue).

**Spin Transport Properties and *PN*-Junction Diodes of MnBi$_2$Te$_4$ Monolayer.** A *pn*-junction diode of MBT-ML (Fig. 2a) is first constructed by using the method of electrostatic doping with *p*- and *n*-type atomic compensation charges,[42] which has been used for various nanodevices modeling.[39,43-45] According to the atomic lattice structures (Fig. 1a), there exist two types of MBT-ML diode structures, i.e., Z-type (along the *x* axis, zigzag at edges) and A-type (along the *y* axis, armchair along the edges). We consider three levels of doping densities of *p*- and *n*-type carriers, i.e., $3 \times 10^{12}$ cm$^{-2}$ (low), $3 \times 10^{13}$ cm$^{-2}$ (medium), and $3 \times 10^{14}$ cm$^{-2}$ (high). They correspond to the typical doping concentration range [$10^{19}$ – $10^{21}$ cm$^{-3}$] in bulk.[42,46] The results for the medium doping concentration are presented here, and results for low and high doping concentrations are shown in the Supporting Information (see Supplementary Fig. 2).

Each diode consists of the drain (D) and source (S) electrodes and the central scattering region with the *pn*-junction. For transport calculations, the D/S electrodes are described by a supercell of *p*- and *n*-doped MBT-ML, whose length is semi-infinite along the transport direction. A forward D-S bias $V_b$ generates a positive current from the D electrode to the S electrode, and vice versa. The spin-resolved electron currents through the *pn*-junction diodes are obtained by[47]

$$I_\sigma(V_b) = \frac{e}{h} \int_{\mu_D}^{\mu_S} T^\sigma(E, V_b)[f_D(E - \mu_D) - f_S(E - \mu_S)] dE, \quad (1)$$

where $\sigma$ indicates the index of spin-up (↑) and -dn (↓) states, and the total current *I* is the sum of $I_\sigma$. *e* and *h* refer to the electron charge and the Planck's constant, respectively. $T^\sigma(E, V_b)$ is the spin-resolved transmission coefficient of the *pn*-junctions. $f_{D(S)} = \{1 + \exp[(E - \mu_{D(S)})/k_B T_{D(S)}]\}^{-1}$ is the Fermi-Dirac distribution function of the D(S) electrode with chemical potential $\mu_{D(S)}$ and temperature $T_{D(S)}$.

Each spin-resolved *I–V* curve of the Z-type *pn*-junction diode of MBT-ML (Fig. 2b) shows a strong unidirectional transport feature, with a high rectifying ratio up to $10^7$ as shown in Fig. 2c (defined as RR = $|I(-V_b)/I(V_b)|$). The current is open in the negative bias side with a low threshold voltage ($V_{on}$ = –0.1 V). In addition, it displays a spin filtering effect. Under a low bias such as –0.1 V, the polarization ratio, defined as PR= $(I_\uparrow - I_\downarrow)/(I_\uparrow + I_\downarrow)$, is 100%. It decreases to a saturation value (21%) as the reverse bias increases. As the current density is close to zero, this MBT-ML based diode is ideal for the use in nanodevice.

Figure 2d shows the spin-up and -down transmission spectra of the Z-type *pn*-junction diode of MBT-ML. The electron transport is easy for reverse biases with large transmission coefficients within the bias window (BW) especially for the spin-up state, but is blocked under a forward bias. When a reverse bias is applied across the *pn*-junction, the bands of the *p*- and *n*-doped terminals shift up and down accordingly, and vice versa. Therefore, compared with the projected local density of states (PLDOS) under zero bias (Fig. 2e), more electronic states of both sides enter the BW, generating a strong electron transmission via Band-to-Band tunneling under a reverse bias, such as –0.8 V (Fig. 2f). In contrast, the electron transmission retains zero under a forward bias due to the band alignment as depicted in Fig. 2g. Furthermore, the PLDOS of spin-up state has smaller gap between valence and conduction band than spin-down state, thus the transmission of spin-up state can do better (Fig. 2f). Within the BW, strong transmission occurs near the Γ point (Fig. 2h). For the spin-up states, transmission coefficients remain high over the whole Brillouin zone, i.e., from Γ to Y (–Y). The transmission becomes zero close to Y (–Y) for the spin-down states. This causes the much reduced transmission for the spin-down states under a reverse bias and appearance of a spin-polarized behavior. These factors result in the strong rectifying and spin filtering effects of the Z-type *pn*-junction diode of MBT-ML. Note that the similar rectifying effect persists as the SOC is included in band and transport calculations (see Supplementary Fig. 1b). Therefore, the concepts we discussed above holds in actual MBT diodes, providing that the reduction of spin mean free path by SOC is insignificant because of the small size of *pn*-junction.

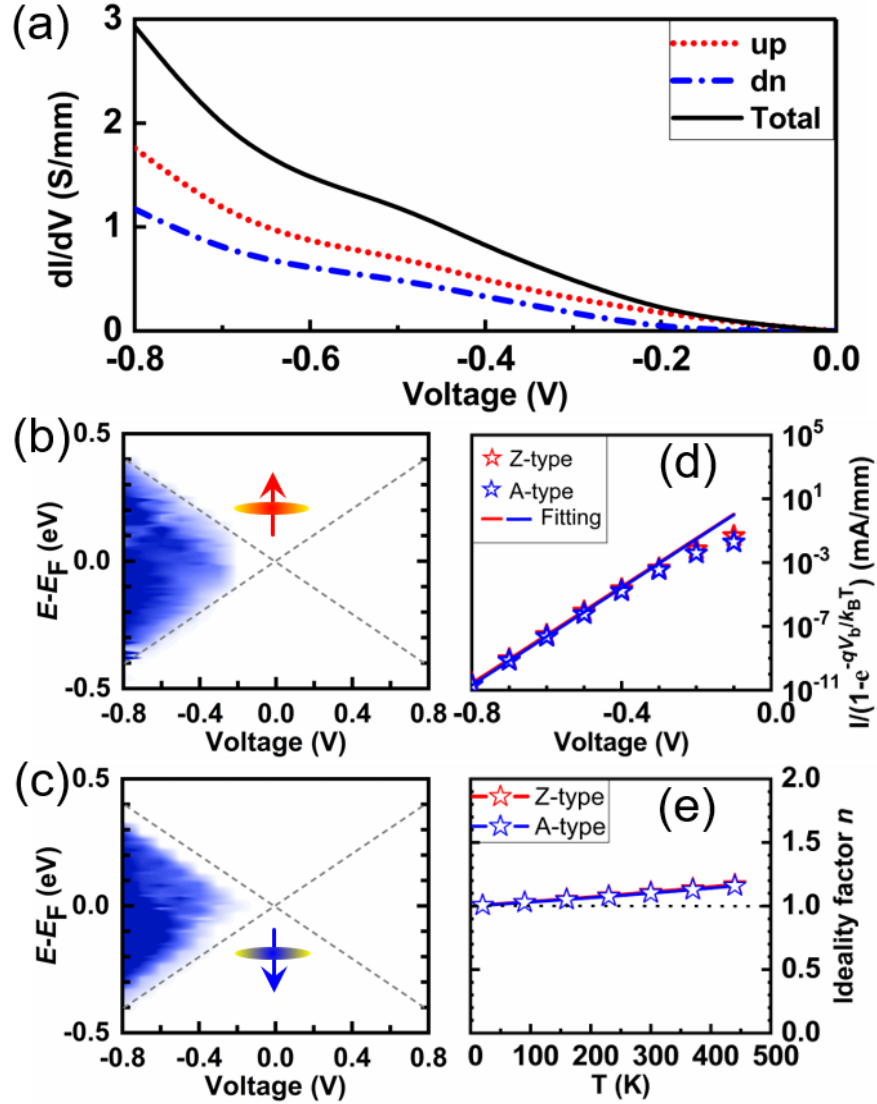

**Fig. 3 Device properties of the *pn*-junction diodes of MnBi$_2$Te$_4$ monolayer. a** Difference conductance curves of Z-type *pn*-junction diodes of MnBi$_2$Te$_4$ monolayer. Spectral currents in the spin-up (**b**) and -down (**c**) states. (**d**) $I/(1 - e^{-qV_b/k_BT})$ against $V_b$ curves. **e** Temperature-dependent ideality factor *n* curves of *pn*-junction diodes of MBT-ML. The dotted line indicates the ideal case (*n* = 1).

The total current and differential conductance (dI/dV) density of the Z-type *pn*-junction diode of MBT-ML are 752 mA/mm (Fig. 2b) and 3 S/mm (Fig. 3a) at –0.8 V, respectively.

The dI/dV curve also shows a spin-polarized behavior, particularly strong under a low bias below −0.2V. The tunneling current mostly appears within the BW, according to the spectral currents shown in Fig. 3b, c. Note that the low doping concentration decreases the current density a little and retains the spin filtering and rectifying effects, while the high doping concentration significantly increases the current density (see Supplementary Fig. 2). The A-type *pn*-junction diode of MBT-ML shows similar rectifying and spin filtering effects with the same mechanisms (see Supplementary Fig. 3, 4) to its Z-type diode. The current along the A direction is slightly smaller than that along the Z direction due to larger carrier effective mass along A direction, implying its weak anisotropy ratio (1.3).[40]

Furthermore, the *I−V* characteristics of the *pn*-junction diodes of MBT-ML can be described by[42]

$$I = I_0 e^{qV_b/nk_BT}(1 - e^{-qV_b/k_BT}), \tag{2}$$

where $q$ and $T$ are the elementary charge and temperature respectively, $I_0$ indicates the saturation current, and $n$ is the so-called ideality factor that is an important parameter to assess how much the *pn*-junction resembles an ideal diode ($n = 1$). One can extract the value of the ideality factor from the slope of a log-plot of $I/(1 - e^{-qV_b/k_BT})$ against $V_b$. As a result, both the ideality factors of both Z- and A-type *pn*-junction diodes of MBT-ML are 1.1 (close to the ideal case) at room temperature (Fig. 3d), and can retain a high value up to 500 K (Fig. 3e).

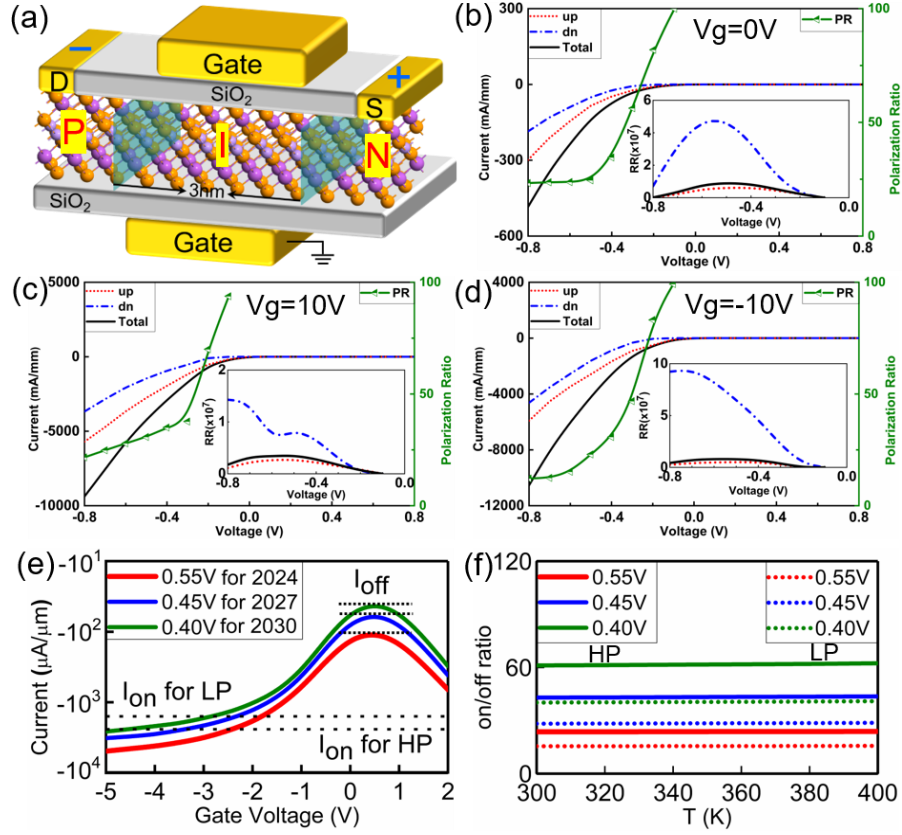

**Fig. 4 Transport properties of the *pin*-junction field-effect transistors of MnBi$_2$Te$_4$ monolayer. a** Schematic of *pin*-junction field-effect transistors of MnBi$_2$Te$_4$ monolayer. **b – d** Bias-dependent current density, rectifying ratio, and polarization ratio curves under the gate voltages of 0, 10, and –10 V. **e** Transfer characteristics at the reverse (S-D) biases of 0.55, 0.45, and 0.40 V, respectively. **f** On/off ratios under the limited temperatures.

**Field-Effect Transistors of MnBi$_2$Te$_4$ Monolayer.** Next, we design the field-effect transistor structures of MBT-ML *pin*-junction (Fig. 4a) and study their spin-dependent field-effect properties. The *pin*-junction is composed of the *p*- and *n*-doped MBT-ML on both sides, and the central intrinsic region (*i*) as the FET channel whose length is 3 nm. Both the top and bottom gates are positioned near the central intrinsic region. Here the spin-resolved electron current through the sub-3-nm *pin*-junction FETs is determined as

$$I_\sigma(V_b, V_g) = \frac{e}{h} \int_{\mu_D}^{\mu_S} T^\sigma(E, V_b, V_g)[f_D(E-\mu_D) - f_S(E-\mu_S)]dE . \qquad (3)$$

Figure 4b shows the *I−V* curves of the Z-type *pin*-junction FET of MBT-ML under zero gate voltage. Compared with the *pn*-junction diode of MBT-ML, the Z-type *pin*-junction FET of MBT-ML shows the same spin filtering and perfect rectifying effects with the same transport mechanisms (see Supplementary Fig. 5), although its current density is slightly suppressed due to the semiconductor nature of the central intrinsic region. Strikingly, when a positive or negative gate voltage is applied, the current density is dramatically enhanced (Figure 4c,d), demonstrating an excellent field-effect behavior.

According to the goals of the ITRS 2015 edition,[48] Figure 4e displays the transfer characteristics of the Z-type *pin*-junction FET of MBT-ML at the reverse (S-D) biases of 0.55 (in the year of 2024), 0.45 (in the year of 2027), and 0.40 V (in the year of 2030), respectively. All of their current densities increase monotonically as the gate voltage increases, and achieve the off-state at $V_g$ = 0.5 V. The corresponding on/off ratios at room temperature for HP(LP) are 24(16), 43(28), and 61(40), respectively. The on/off ratios remain constant up to 400 K (Fig. 4f), and may further increase to achieve the three goals in 2024, 2027, and 2030.

The A-type *pin*-junction FET of MBT-ML shows the same field-effect properties to its Z-type counterpart and has even larger on/off ratios (see Supplementary Fig. 6,7). Besides, we also construct and investigate two other types of MBT-ML FETs, i.e., *pip*- and *nin*-junction FETs (see Supplementary Fig. 8:13). Each of them shows an interesting negative differential resistive (NDR) effect[49] and a spin filtering effect with larger polarization ratios. Figure 5 schematically illustrates their NDR mechanisms and the rectifying mechanism of the *pin*-junction FET that is consistent with the previously discussed *pn*-junction diode. For the *pip*-junction FET, its VB is adjacent to the $E_F$ and plays a critical role in the NDR behavior. Very differently, for the case of *nin*-junction FET, its CB adjoins to the $E_F$ and dominates the electron transport.

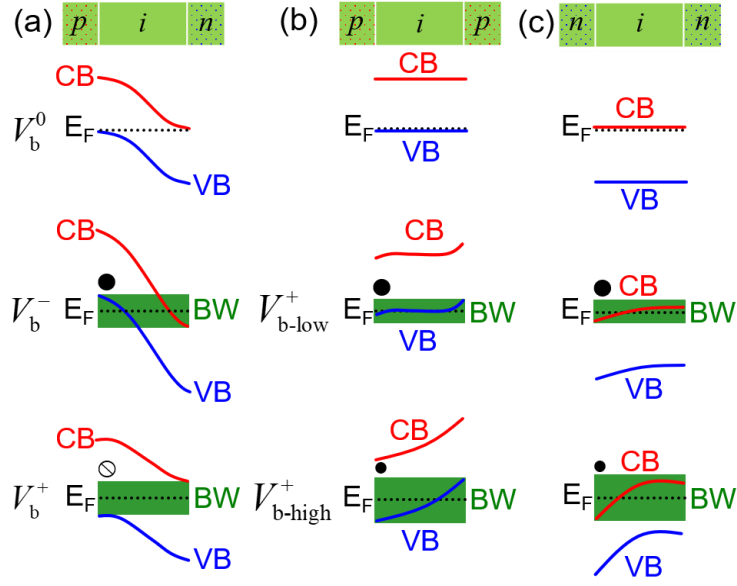

**Fig. 5 Rectifying and NDR mechanisms of the transistors of MnBi$_2$Te$_4$ monolayer. a** Rectifying mechanism of *pin*-junction FETs of MnBi$_2$Te$_4$ monolayer. **b** NDR mechanism of *pip*-junction FETs of MBT-ML. **c** NDR mechanism of *nin*-junction FETs of MBT-ML. Green shadow is the bias window. ● and • refer to the strong and weak transport, and ⊘ indicates the prohibited transport.

**Phototransistors of MnBi$_2$Te$_4$ Monolayer.** We further investigate the photoelectric performance of MBT-ML and its potential applications in spin-photonics. The optical-conductivity σ (|real| part) is found to be almost unchanged as different functionals are used in our calculations, i.e., spin-polarized generalized gradient approximation[50] and local density approximation.[51] The photoconduction process is opened as the photon energy is higher than the energy gap of MBT-ML (Fig. 1c,d). It has a broad σ peak in the whole visible region (Fig. 6a) for both the spin-up and -down states, and it is therefore promising for developing photovoltaic devices within the AM1.5 standard.[52] Whereafter, a phototransistor based on the *pin*-junction of MBT-ML is designed (Fig. 6b) and its photoelectric transport properties under illumination are unveiled, as well as the regulation effect from the gate electrode.

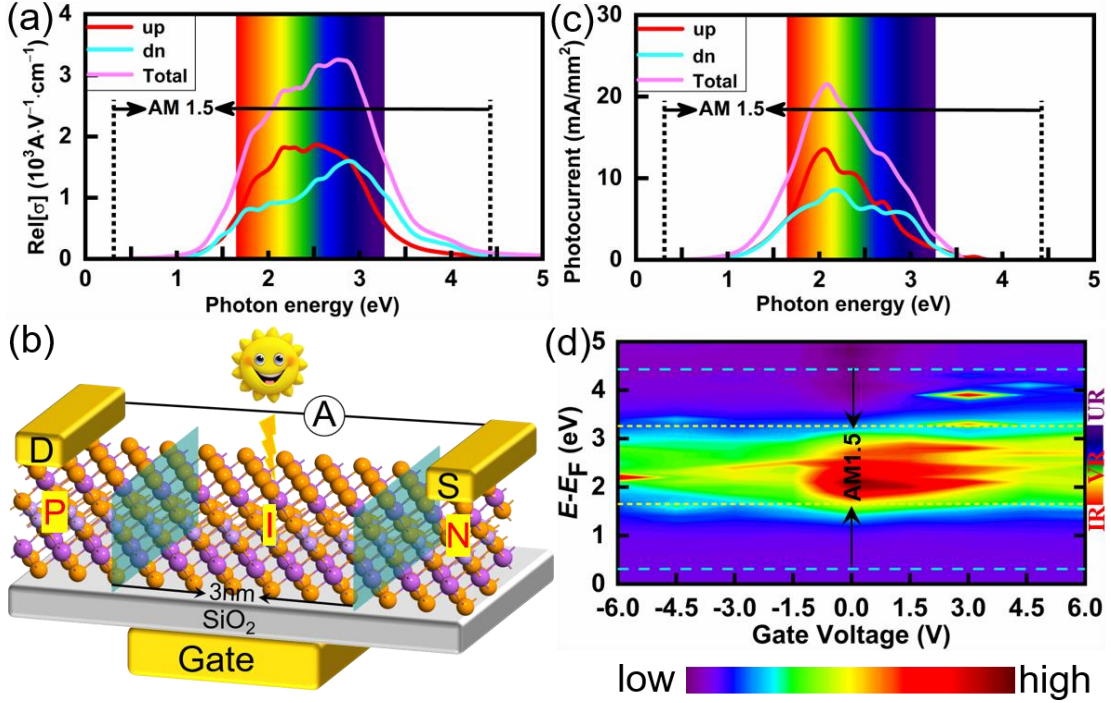

**Fig. 6 Photoelectric properties of the MnBi$_2$Te$_4$ monolayers. a** Optical-conductivity of MnBi$_2$Te$_4$ monolayer. The embedded spectrum pattern displays the visible region. **b** Schematic of the *pin*-junction phototransistor of MBT-ML. **c** Spin-resolved photocurrent density of Z-type *pin*-junction phototransistor of MBT-ML under zero bias (without power). **d** Gate-dependent photocurrent spectra of Z-type *pin*-junction phototransistor of MBT-ML under zero bias. IR, VR, and UR refer to the infrared, visible, and ultraviolet region, respectively.

In this work, linearly polarized light is considered and the incident photon energy is set from 0 to 5 eV. The first-order correction to the photo-generated current into electrode $\alpha$ = D/S due to absorption of photons with frequency $\omega$ is given by[44,53]

$$I_\alpha = \frac{e}{h} \int_{-\infty}^{\infty} \sum_{\beta=D,S} [1 - f_\alpha(E)] f_\beta(E - \hbar\omega) T^-_{\alpha,\beta}(E) - f_\alpha(E)[1 - f_\beta(E + \hbar\omega)] T^+_{\alpha,\beta}(E) dE . \quad (4)$$

The total photocurrent is obtained by $I_{ph} = I_D - I_S$. More details about the photocurrent are provided in the Supplementary D1. Under zero bias (without power), the Z-type *pin*-junction phototransistor of MBT-ML has a good photo response in the visible region due to its large σ in that region, and its total photocurrent is as high as 22 mA/mm$^2$ (Fig. 6c), much larger than a silicon solar-cell device.[44] In addition, the spin-up photocurrents have a larger proportion and a stronger peak in yellow light region, demonstrating its key role in photoelectric sensors.

Table I. Device functions of various MnBi$_2$Te$_4$ monolayer transistors.

| Transistors | Rectifier | Spin filter | NDR device | Photoelectric sensor | Photovoltaic device |
|---|---|---|---|---|---|
| *pn*-junction diode | ✓ | ✓ | | | |
| *pin*-junction FET | ✓ | ✓ | | | |
| *pip*-junction FET | | ✓ | ✓ | | |
| *nin*-junction FET | | ✓ | ✓ | | |
| *pin*-junction phototransistor | | | | ✓ | ✓ |
| *pip*-junction phototransistor | | | | ✓ | |
| *nin*-junction phototransistor | | | | ✓ | |

Generally, a gate electrode can be employed to tune the optical response of a phototransistor.[54,55] The applied gate voltages significantly influence the photoelectric performance of the Z-type *pin*-junction MBT-ML phototransistor. For instance, it is easy to generate a strong photocurrent peak under low gate voltages (Fig. 6d), while high gate voltages lower its photoelectric performance and should be avoided in a photovoltaic device. Therefore, the Z-type *pin*-junction phototransistors of MBT-ML can be utilized as photovoltaic devices or photoelectric sensors for detecting yellow light. For the A-type *pin*-junction phototransistor of MBT-ML, whose spin-down state plays a key role, it generates a peak in the yellow (for spin-up state) and the blue light (for spin-down state), respectively. The total photocurrent amplitude is half of the Z-type phototransistor (see Supplementary Fig. 14). Last, the photoelectric performances of other types of MBT-ML (i.e., *pip*- and *nin*-junction) are lower than its *pin*-junction phototransistors (see Supplementary Fig. 15:18), and they may be overshadowed in photovoltaic devices.

In summary, we have designed various conceptual nanodevices based on the MnBi$_2$Te$_4$ monolayer (listed in Table I) and investigated their spin-dependent transport properties through extensive first-principles calculations. Both the Z-type and A-type *pn*-junction diodes of MBT-ML show a spin filtering effect with a high spin polarization ratio under a low bias and a strong rectifying effect with a high rectifying ratio (up to the magnitude of $10^7$). Even though most our discussions were based on results without spin-orbital coupling for the easiness of spin separation, calculations with the inclusion of SOC indicate that all conclusions hold, at least at the qualitative level. The MBT-ML *pn*-junctions have an excellent ideality factor (close to 1) insensitive to temperature in a reasonable range. The sub-3-nm *pin*-junction FETs of MBT-ML exhibit the same rectifying and spin filtering effects. The gate voltages may further significantly enhance the current through the FETs and achieve an on-off action. The *pip*- and *nin*-junction FETs of MBT-ML also show an interesting NDR and excellent spin filtering effects. Moreover, the MBT-ML has large optical conductivity in the visible region and the Z-type *pin*-junction phototransistor has the largest response to yellow light which can be used in optoelectronic devices. Overall, our results indicate that the MnBi$_2$Te$_4$ monolayer is a multifunctional material and holds promise for developing a verity of next-generation nanodevices with applications in spintronics and photonics.

## METHODS

All the first-principles self-consistent calculations are performed by the density-functional theory combined with non-equilibrium Green's function method using the Atomistix Toolkit code.[56-58] The Norm-Conserving pseudopotentials,[59] linear combinations of atomic orbitals basis sets, and the spin-polarized Perdew-Burke-Ernzerhof exchange-correlation functional at the level of generalized gradient approximation (GGA) are used.[50,60] The GGA+U method is employed to describe the localized 3d orbitals of Mn atoms and U = 4 eV is adopted according to previous tests.[41] A real-space grid density mesh cut-off of 100 Ha is used. The Monkhorst-Pack $k$-point grids $1 \times 5 \times 200$ and $1 \times 9 \times 150$ are used to sample the Brillouin zone for the electrodes of Z-type and A-type MBT-ML devices, respectively. The total energy tolerance and residual force on each atom are less than $10^{-6}$ eV and 0.001 eV/Å in the structural relaxation of both lattice constants and atomic positions, respectively.

## DATA AVAILABILITY

The data that support the findings of this study are available from the corresponding author upon reasonable request.

## ACKNOWLEDGMENTS

We acknowledge funding from the National Natural Science Foundation of China (Nos. 11774079 and 61774059), the Scientific and Technological Innovation Program of Henan Province's Universities (No. 20HASTIT026), the Natural Science Foundation of Henan (No. 202300410226), the Natural Science Foundation of Henan Normal University (No. 2020PL15), the Henan Overseas Expertise Introduction Center for Discipline Innovation (No. CXJD2019005), and the HPCC of HNU. R. Wu acknowledges funding from the US DOE-BES (No. DE-FG02-05ER46237). We thank F. Xue at Tsinghua University, W. Ju and D. Kang at HNUST for helpful discussions.

## AUTHOR CONTRIBUTIONS

Y.A., T.W. and H.W. performed the most calculations in this paper. Y.A. wrote this manuscript. R.W., W.L., K.W. and Y.H. gave many significant suggestions and help to revise this manuscript. S.G, C.M, M.Z, C.Z. and S.M gave many significant suggestions. All the authors contributed to the discussion and preparation of the manuscript.

## COMPETING INTERESTS

The authors declare no competing interests.

## ADDITIONAL INFORMATION

Supplementary information is available for this paper at XXX

**Correspondence** and requests for materials should be addressed to Y.A.

**Reprints and permission information** is available at http://www.nature.com/reprints

**Publisher's note** Springer Nature remains neutral with regard to jurisdictional claims in published maps and institutional affiliations.

simple. *Phys. Rev. Lett.* **77**, 3865–3868 (1996).